\documentstyle[11pt,paspconf,psfig,twoside]{article}

\begin{document}

\title{RX J2115-5840: confirmation of a new near-synchronous Polar}

\author{Gavin Ramsay,$^{1}$ David Buckley,$^{2}$ Mark Cropper,$^{1}$
M. Kate Harrop-Allin$^{1}$}
\affil{$^{1}$Mullard Space Science Lab, University College London,
Holmbury St. Mary, Dorking, Surrey, RH5 6NT, UK\\
$^{2}$SAAO, PO Box 9, Observatory 7935, Cape Town, South Africa}

\begin{abstract}
Schwope et al (1997) suggested that the newly discovered Polar RX
J2115-5840 is a near-synchronous system. We have obtained circular
polarisation observations of RX J2115-5840 which show that the spin
and orbital periods differ by 1.2\%. We find the first direct evidence
of `pole-switching' in a near-synchronous Polar. Further our data
requires that the accretion flow must be directed onto the same
magnetic field line at all spin-orbit beat phases implying that at
some phases the flow must follow a path around the white dwarf before
accreting.
\end{abstract}


\section{Introduction}

RX J2115-5840 (EUVE 2116--58.6) was discovered during the {\sl ROSAT}
(Voges et al 1996) and {\sl EUVE} (Bowyer et al 1996) all sky surveys.
Subsequent ground based observations confirmed the source as a
Polar. Schwope et al (1997) suggested that the orbital and spin
periods of RX J2115-5840 differ by $\sim$1\%. This would make it the
fourth near synchronous Polar and the first one below the 2--3 hr
period gap. We obtained 2 weeks of white light polarimetric
observations in July-Aug 1997 at SAAO to investigate this possibility.

\section{Results}

\begin{figure}
\begin{center}
\setlength{\unitlength}{1cm}
\begin{picture}(6,8)
\put(-3,-2.){\includegraphics{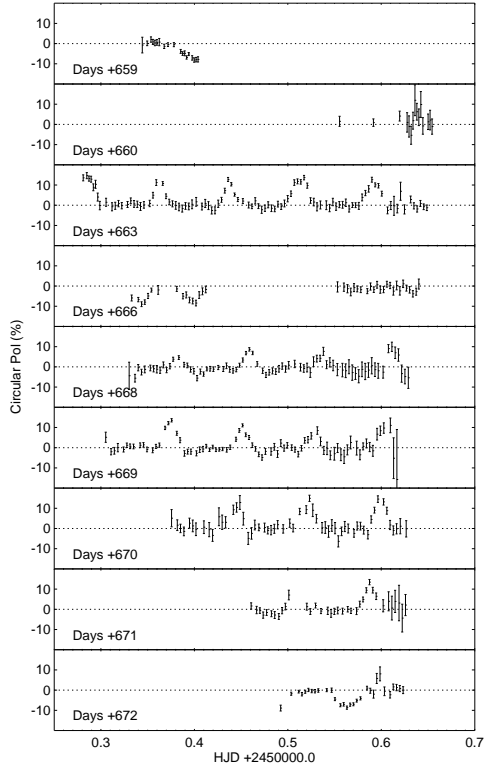}}
\end{picture}
\end{center}
\caption{The circular polarisation data. The HJD is shown on the left
hand side +2450000.0.}
\label{cpol} 
\end{figure}

The circular polarisation data are shown in Fig. \ref{cpol}: the
circular polarisation is generally either close to zero or shows
positive excursions. However, there are occasions when negative
polarisation is seen (HJD 2450000+ 659, 666 and 672).  These
observations suggests that RX J2115--58 is not fully synchronised.

\begin{figure}
\begin{center}
\setlength{\unitlength}{1cm}
\begin{picture}(8,7)
\put(-4,-28.5){\includegraphics{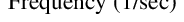}}
\end{picture}
\end{center}
\caption{From the top: the amplitude spectrum of the circular polarisation
data; the amplitude spectrum in greater
detail over three frequency intervals; the window function and the 
amplitude spectrum pre-whitened using the frequencies mentioned in the text.}
\label{power} 
\end{figure}

To investigate the circular polarisation data more closely, we used a
Discrete Fourier Transform to obtain an amplitude spectrum (Fig
\ref{power}).  The highest amplitude peak corresponds to a period of
110.889 mins -- similar to the spectroscopic period of 110.8 mins
reported by Vennes et al (1996). The second highest peak corresponds
to a period of 109.547 mins -- similar to the shorter of the two
possible optical photometric periods, 109.84 and 109.65 mins, reported
by Schwope et al (1997). If we assume that the binary orbital period
is $\Omega$=110.889 min and the spin period of the white dwarf is
$\omega$=109.547 min we find the following more complex frequencies in
the amplitude spectrum: 3$\Omega$, 2$\omega$, 3$\omega$,
$\omega\pm\Omega$, 2$\omega+\Omega$, 4$\Omega$--3$\omega$. Any other
frequencies which are present have amplitudes lower than 1\%. The
spin-orbit beat frequency corresponds to a period of 7.1 days.

\section{The spin-orbital beat period}

To make a more detailed investigation of these data we folded the
circular polarimetry on the proposed spin and orbital periods and the
spin-orbital beat period (the left hand panel of Fig
\ref{beat_phase}). Folding the data which correspond to a discrete
beat phase on the proposed spin period of the white dwarf, we find
that the polarisation curve shows a negative excursion lasting
approximately half the spin cycle. At other spin phases the
polarisation is close to zero (the right hand panel of Fig
\ref{beat_phase}). At $\phi_{(\omega-\Omega)}$=0.20 the polarisation
is not significantly modulated. At other beat phases a prominent
positive hump is seen in the folded spin polarisation curves, the peak
of which advances in phase as $\phi_{(\omega-\Omega)}$ increases.

\begin{figure}
\begin{center}
\setlength{\unitlength}{1cm}
\begin{picture}(8,7)
\put(-3.5,-29.5){\includegraphics{rx2115_fig3.ps}}
\end{picture}
\end{center}
\caption{Left panel: The circular polarisation folded on the spin
period (top panel); folded on the orbital period (middle panel),
folded on the orbital-spin beat period, $\omega-\Omega$. Phase zero
was arbitrarily chosen. Right panel: The circular polarisation data
folded and binned on the proposed spin period of the white dwarf, as a
function of the spin-orbital beat period. The beat phase is shown at
the right hand side.}
\label{beat_phase} 
\end{figure}

\section{Pole-Switching}

In fully synchronous Polars, the accretion flow is locked with respect
to the binary orbital rotation frame and the bulk of the accretion
flow is thought to be directed onto the geometrically preferred
magnetic pole of the white dwarf. However, in the case of
near-synchronous Polars, the accretion flow rotates around the
magnetic field of the white dwarf on the spin-orbit beat period. This
has the effect that the accretion flow will be directed preferentially
onto first one then the other magnetic pole of the white dwarf. At two
phases of the spin-orbit beat period we expect that the flow will be
equally directed onto both poles. This `pole-switch' will manifest
itself most obviously in the circular polarisation curves where the
polarisation will change sign after the accretion flow has `switched'
poles. This is seen in the right hand panel of Fig \ref{beat_phase}
where at $\phi_{(\omega-\Omega)}\sim$0.00 the polarisation is
modulated with a positive hump, but at
$\phi_{(\omega-\Omega)}\sim$0.07 and 0.17 it is modulated with a
negative hump.

\begin{figure}
\begin{center}
\setlength{\unitlength}{1cm}
\begin{picture}(8,4)
\put(-2,8.5){\includegraphics{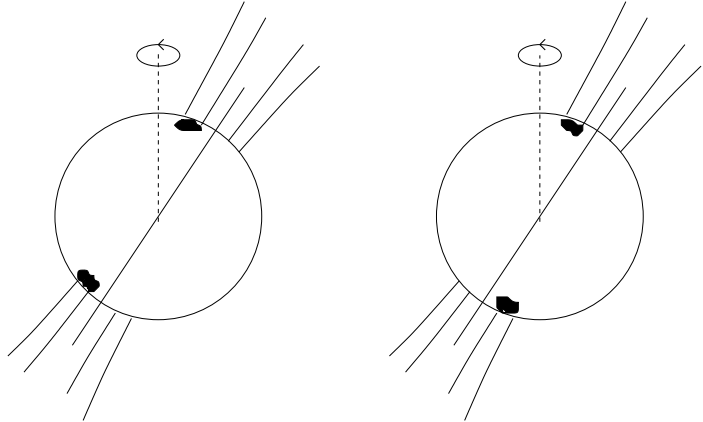}}
\end{picture}
\end{center}
\caption{Our two accretion scenarios. In the left hand panel the
accretion flow is directed onto one or other footprint of the same set
of magnetic field lines at all spin-orbit beat phases. In the right
hand panel the flow is directed onto a diametrically opposite set of
field lines at different beat phases. The spin axis of the white dwarf
is shown as a dotted vertical line. We favour the scenario described
in the left hand diagram.}
\label{scenario} 
\end{figure}

We consider two accretion scenarios which are described in Fig
\ref{scenario}. The phasing of the data on the spin, orbital and beat
phases requires that the accretion flow must be directed onto the same
magnetic field line at all spin-orbit beat phases implying that at
some phases the flow must follow a path around the white dwarf before
accreting. This is difficult to reconcile with simple views of how the
accretion stream attaches onto the magnetic field of the white dwarf.
Possible reasons for this are described in a more detailed paper that
has been submitted to {\sl MNRAS}.


\begin{references}
\reference Bowyer, S., et al, 1996, ApJS, 102, 129
\reference Schwope, A. D., et al, 1997, AA, 326, 195
\reference Vennes, S., et al, 1996, AJ, 112, 2254
\reference Voges, W., et al, 1996, IAUC 6420
\end{references}
\end{document}